\documentstyle[preprint,aps]{revtex}
\begin{document}
\draft

\title{Threshold Laws for the Break-up of Atomic
Particles into Several Charged Fragments }

\author{M.~Yu.~Kuchiev and V.~N.~Ostrovsky\cite{SPb}}

\address{School of Physics, University of New South Wales,
Sydney 2052, Australia}

\maketitle

\begin{abstract}

The processes with three or more charged particles in the
final state exhibit particular threshold behavior, as inferred by
the famous Wannier law for (2$e$ + ion) system.
We formulate a general solution which  determines the threshold
behavior of the cross section
for multiple fragmentation.
Applications to several  systems of particular
importance with three, four and five leptons (electrons and positrons)
in the field of charged core; and two pairs of identical particles with
opposite charges are presented.
New threshold exponents for these systems are predicted,
while some previously suggested threshold laws are revised.

\end{abstract}



\section{INTRODUCTION}
\label{intro}
The famous Wannier \cite{Wannier} threshold law has quite an unusual
status among other threshold laws in quantum mechanics. Being based
on an appealing mechanism, it has inspired a  large number of studies
where the law was rederived, extended, tested or rebutted.
The intensity of these studies does not show a decrease with time
as testify some representative references to the recent
publications \cite{Wat91} \cite{Hall} \cite{KO92}
\cite{Lub} \cite{KOWang} \cite{1/4} \cite{EWRM} \cite{MO}
\cite{Sim} \cite{KOsharing} \cite{KOZPhys} \cite{Posexp}
\cite{Wat} \cite{Temkin-Poet} \cite{KOPCI} \cite{Kuch}
\cite{Macekpos} \cite{Fantip} \cite{Rost} \cite{Grucom}.
A complete bibliography on the subject would be immense.

In this paper we suggest a  method which
generalizes
the Wannier mechanism when break-up of the quantum system on a
{\it large number}\/ (four or more) of charged fragments is concerned.
Apparently, for the first time the particular case of the problem
was treated  in 1976 in the important paper by
Klar and Schlecht \cite{KS} where  the threshold
law was derived
for escape of three electrons from the charged core. It was
suggested that the receding electrons form a symmetrical configuration
of equilateral triangle with the positively charged core in its center.
The treatment was quite involving and specialized being based on
hyperspherical coordinate system \cite{ftnt1}.
Later Gruji\'{c} \cite{Gru3e} rederived the same result using
the standard Cartesian coordinates where the symmetry considerations
are easy to apply explicitly in full extent. Gruji\'{c} considered
also
some other systems along the same lines 
\cite{Gru2ep}\cite{Gru4e}\cite{Gru5e}
(see more details in Sec.\ref{parti}).
The threshold law for the three-electron escape seems to find
support in the experimental data on the near-threshold double
ionization of atoms by electron impact \cite{exp3e}. Later
Feagin and Filipczyk \cite{FF} claimed an existence of a
complementary law which is manifested at energies somewhat
above a threshold, see critical discussion
in Section \ref{parti}A.

The interest to the problem was renewed recently when
two electrons and positron receding from the core with $Z=1$ charge
were considered by Poelstra {\it et al}\/ \cite{Poel}.
A brief note by Stevens and Feagin \cite{FBul} on complete
fragmentation of H$_2$ molecule is  also to  be mentioned.
The final state in the reaction with a positron
could be produced by double ionization
of a negative ion by positron impact. However, the forthcoming
experiments by Knudsen and co-workers \cite{Knudsen} concern
positron impact double ionization of neutral atoms ($Z=2$) where
the threshold law has not yet been available.
Its derivation was one of
motivations for the present study.  Eventually it has
developed into a general approach to the multi-fragmentation
problem which possesses  two important advantages.
Firstly, our method describes a general situation
with arbitrary number of charged fragments in simple terms
in an arbitrary coordinate frame.
Secondly, it is convenient and reliable for practical realizations.
This allows us to clarify important conceptual aspects of the problem
which  were  misunderstood or misinterpreted previously.
Comparing our solution
with known in literature results we reproduce a number of
threshold  exponents for different systems.
At the same time we find that several published previously results
need improvement, in particular, we
revise the threshold law for $2e^- + e^+$ escape.
A general nature of the developed method
is illustrated by consideration of new  complicated situations
with up to six charged particles in the final state where a number
of new threshold exponents is predicted.

In Section II we introduce particular configurations
which will be called {\it scaling configurations}.
They describe a multidimensional dynamic potential saddle,
generalizing the {\it Wannier ridge}\/ which is well known for
(2$e$ + ion) system.
These configurations are related to {\it rectilinear trajectories of all
particles} in the system and play a crucial role for the complete
fragmentation process close to its threshold.
SCs embrace the essence of previous treatments of particular
systems, but avoid  attachment to some special
theoretical formalism and related technical complications.
The closest analogue of our general approach in particular case
of three-body Coulomb systems could be found
in papers by Simonovi\'{c} and Gruji\'{c} \cite{Grumass}.

Description of  small deviations from SC is
given in terms of a set of harmonic oscillators and inverted
oscillators (Section \ref{small}). The later ones describe  unstable modes
which  govern the  threshold law.
They are quantized following a general scheme
suggested by Kazansky and Ostrovsky \cite{KOJETP} \cite{KOharm}.
This allows us to construct a reduced form of the wave function for
the system of charged particles and derive the threshold law
in Section IV generalizing a procedure
used previously by Kazansky and Ostrovsky \cite{KO92} \cite{KOZPhys}
for derivation of the conventional Wannier law.
Application of a developed
general scheme to some particular systems (Section V)
is followed by concluding discussion of special features of
the Wannier-type threshold laws, with an emphasis on a relation
between the underlying statistical and dynamical aspects
of the problem (Section VI).

\section{SCALING EXPANSION}
\label{QEEC}
Our goal is to consider some atomic  process which breaks
an atomic particle
into  several charged fragments  for low excess energy $E$.
In this situation the  motion of the fragments
in the final state of the reaction
can be described in the semiclassical approximation because
a typical variation of the Coulomb potential $U_C\sim 1/r$ on the
wavelengths of the fragments $\lambda \sim 1/\sqrt {M E}$,
$\delta U \simeq \lambda /r^2$,
is less than a typical kinetic energy $T \sim E$
\[\delta U \ll T \]
inside the Coulomb zone $r \le r_C = 1/E$ where the major events take
place.
Therefore  the first thing to do is to find classical trajectories
which lead to the desired final state with total fragmentation.

It is very important that for low energy $E$
there exists a severe
restriction onto  these trajectories.
To see this let us imagine what is happening with distances
separating fragments when they move out of the reaction domain.
If a distance separating some pair of two attracting
fragments diminishes with time, then one should expect
that this pair of fragments
can be considered as a dipole which interacts with the rest fragments.
This interaction can transfer the kinetic energy of the
two fragments to the other fragments.
Therefore one has to expect that eventually these two fragments
will loose enough energy and form a bound state.
If this event happens then the desired {\it total}
fragmentation is not achieved.
This discussion shows that one should  look for  those trajectories
which exhibit a  monotonic increase of distances
separating  the  fragments.
The point is that the lower is the available
above-threshold energy the more restrictive this condition is.

It is convenient to present the discussed situation considering
the potential energy in the multidimensional configuration
space where its behavior can be described
as  kind of ``valleys separated by ridges''. This physical picture
first suggested by Wannier for a  particular class
of reactions was discussed by
Fano \cite{Fano,FN} in general case.
If a system occupies some place
on some ridge  then its trajectory  can either go down
into some valley  where
a bound state of some fragments is created,
or continue to propagate along the ridge.
For the total
fragmentation one should find a
classical trajectory along a top of some
ridge which leads from the region
of small separation of fragments into
the final state with infinite separation.
It is clear that
the lower is the above-threshold energy, the closer
a trajectory should be to the top of the ridge.

Generally speaking there might exist several such ridges
which lead to the final state with total fragmentation.
In this work we study a particular ridge, which will be called
the {\it  scaling configuration} (SC).
For all the systems considered up to now we have
found that this configuration exists.
More than that, for a given system there may exist
several different SCs. A number of examples demonstrating
this property is considered in  Section \ref{parti}.
Therefore one might suspect that SC is a general
feature, though this latter statement is not
verified so far.
For two electrons in the field of an ion
this ridge coincides with the Wannier ridge.

The basic idea is simple. As was said above one has to be sure
that distances separating  fragments monotonically increase with time.
This condition is definitely satisfied
if a trajectory  describing $N$ particles which have masses
$m_j~(j = 1,2,\cdots N) $
obeys the following conditions
\begin{equation} \label{scale}
\vec{r}_j(t) = \phi(t) \, \vec{\rho}_j ,~~~~~j = 1,2,\cdots N,
\end{equation}
which are valid in the center of mass reference frame
$\sum_j m_j \vec r_j(t) = 0.$
We
shall
refer to a trajectory satisfying Eq.(\ref{scale}) as SC.
The time-independent vectors $ \vec \rho_j $
describe the shape of the SC, while
the function $\phi(t)$ gives the overall scaling factor.
We will see below that this function monotonically increases
in time thus ensuring that all distances increase as well.
Therefore
this type of motion definitely
results in total fragmentation
avoiding traps into  potential valleys.
It is convenient to normalize the scaling function
to unity  for some initial
moment of time  $t_0$
\begin{equation}\label{phi1}
\phi(t_0) =1.
\end{equation}
For  this normalization
the vectors $\vec \rho_j$ play a role of coordinates
of the particles at this initial  moment of time
$\vec r_j(t_0) = \vec \rho_j$
in the center of mass reference frame
\begin{equation}\label{orig}
\sum_j m_j \vec \rho_j = 0.
\end{equation}
Notice that in SC all degrees of freedom except the only one
describing the overall scaling factor are frozen. In this sense
the SC describes a {\it quasiequilibrium } of the system.

It is obvious that to satisfy (\ref{scale}) one should
choose appropriately the initial coordinates $\vec \rho_j$.
Let us formulate restrictions on them. Notice first of all that
in the SC the accelerations of the particles are
\begin{equation} \label{d2t}
\frac{d^2 \vec{r}_j(t)}{dt^2} = \frac{ d^2\phi(t)}{dt^2} \, \vec{\rho}_j.
\end{equation}
We presume purely Coulomb interaction,
or consider Coulomb asymptote in more complicated cases,
which is possible
because the important distances are large ($r \sim r_C = 1/E$) for
low above-threshold energy. Therefore the potential energy of the system
of $N$ fragments is
\begin{eqnarray}
U = \sum_{m>n} \frac{q_m q_n}{\left| \vec{r}_m - \vec{r}_n \right|} .
\end{eqnarray}
Here $q_j$ is a charge of a $j$-th fragment. The forces
$\vec{F}_j$ for SC are time-scaled as
\begin{eqnarray} \label{f}
\vec{F}_j(t) = -  \frac{\partial U}{\partial
\vec{r}_j}
=\frac{1}{\phi(t)^2}
\sum_{n\ne j} q_j q_n \frac{\vec \rho_{jn} }{ \rho_{jn}^3 } ,
\end{eqnarray}
where $\vec \rho_{jn} = \vec \rho_{j} -\vec \rho_{n}$.
Substituting (\ref{d2t}),(\ref{f}) in the Newton equation of motion
one finds the following relation
\begin{equation}\label{2Nl}
\frac{ d^2 \phi(t)}{dt^2} m_j \vec \rho_j =
\frac{1}{\phi(t)^2}
\sum_{n\ne j} q_j q_n \frac{\vec \rho_{in} }{ \rho_{in}^2 }.
\end{equation}
It is easy to see that it can be satisfied only if
two conditions are fulfilled. Firstly, the scaling function
should satisfy an equation
\begin{equation}\label{alpha}
\frac{d^2\phi(t)}{dt^2} = -\frac{\alpha}{\phi(t)^2},
\end{equation}
where $\alpha > 0$ is a time-independent constant which is discussed
in detail below. One obviously recognizes in (\ref{alpha}) the equation
describing a one-dimensional motion of a particle
with unite mass and unite charge
in the attractive Coulomb field created by the  charge $\alpha$.
Secondly, the validity of (\ref{2Nl}) needs that
the  vectors
$\vec \rho_j$ satisfy the following system of equations
\begin{equation} \label{QEC}
 \alpha \vec \rho_j = \vec a_j \equiv -\frac{1}{m_j}
\sum_{k \ne j}q_j q_k\frac{\vec \rho_{jk} }{ \rho_{jk}^3},
\end{equation}
They state that accelerations of each fragment $\vec a_j$ is
proportional to its coordinate vector at the initial moment
of time.
Eqs.(\ref{QEC}) are shown to arise as  conditions which are necessary
for existence of SC. It is easy to see that they provide
sufficient conditions as well.
To verify this statement let us assume that we have a solution
of (\ref{QEC}). Then we can consider a trajectory
with the following initial conditions. Firstly, we can choose
initial coordinates as $\vec r_j(t_0) = \vec \rho_j$.
Secondly, we can always choose initial velocities
be proportional to  coordinates
\begin{equation}\label{vel}
\frac{d \vec r_j(t_0)}{dt} = \beta r_j(t_0),
\end{equation}
where  $\beta$ is some positive constant which depends
on the energy, $\beta \sim \sqrt E$.
 From (\ref{QEC}) we find that accelerations at the initial
moment of time are also proportional to coordinates
\begin{equation} \label{accel}
\frac{d^2 \vec r_j(t_0)}{dt^2} = \alpha  \vec r_j(t_0).
\end{equation}
Thus for the considered trajectory
both the velocities and accelerations
linearly depend on coordinates
at the initial moment of time.
Combining this fact with the Newton equations of motion
we conclude that  the velocities (and accelerations)
remain to be proportional to
the coordinates for any moment of time
\begin{equation} \label{betat}
\frac{ d \vec r_j(t)}{dt} = \beta(t) \vec r_j(t).
\end{equation}
Here $\beta(t)$ is some positive function, $\beta(t_0) = \beta$.
Integrating (\ref{betat})
we conclude that the time
variation of distances  does
exhibit  scaling condition (\ref{scale}), in which
$\dot \phi (t) = \beta(t)$.

This discussion shows that the SC (\ref{scale}) exists if
and only if  Eqs.(\ref{QEC}) are satisfied.
There are  $N$ vector variables
$\vec \rho_j,~j=1,2,\cdots N$ and one scalar variable
$\alpha$ in these equations.
Obviously not all of them are independent because
there are seven transformations which do not change the
given SC. Three of them  correspond to shifts of the
SC center of mass. Three others describe rotations of the
SC as a whole. One more transformation describes
the overall scaling of SC
\begin{eqnarray}\label{scaler}
\vec \rho_j \rightarrow \vec \rho_j'&=&  \lambda \vec \rho_j,~~~j=1, 2,
\cdots N, \\ \label{scalea}
\alpha \rightarrow \alpha' &=& \lambda^{-3} \alpha,
\end{eqnarray}
with $\lambda >0$. According to Eq.(\ref{alpha})
the scaling of $\alpha$ (\ref{scalea})
should be accompanied by a
corresponding scaling of $\phi(t)$, namely
$\phi(t)\rightarrow \phi'(t) = \lambda^{-1} \phi(t)$.
Notice that the latter transformation can be interpreted
as a shift of the initial moment of time
\begin{equation}\label{time}
t_0 \rightarrow t_0',
\end{equation}
where according to Eq.(\ref{phi1}) $t_0'$ should satisfy
\begin{equation}\label{t0'}
\phi'(t_0') = \lambda^{-1} \phi(t_0') = 1.
\end{equation}
It is easy to see that Eqs.(\ref{QEC}) remain invariant under
the discussed above seven transformations,
i.e. the shifts, rotations and  scaling, allowing one
to consider them as a set of $3N-7$ equations for
$3N-7$ independent variables.
When solving these equations it is convenient to treat $\alpha$
as a constant parameter which governs the overall scale and can
be chosen arbitrary (for example $\alpha = 1$).

At SC the system Hamiltonian
\begin{equation} \label{hami}
H = \sum_{j=1}^N \frac{\vec{p}_j^{\, 2}}{2 m_j} + U ,
\quad \quad \quad
\vec{p}_j = m_j \frac{d \vec{r}_j}{dt}
\end{equation}
is reduced to
\begin{equation} \label{Hamphi}
H_0 = \frac{1}{2} \, {\cal M} \left( \frac{d \phi}{dt} \right)^2
- \frac{{\cal Q}_0}{\phi} ,
\end{equation}
where
\begin{eqnarray} \label{M}
{\cal M}& =&  \sum_{j=1}^N m_j \vec{\rho}_j^{\, 2} , \\  \label{Q}
{\cal Q}_0 & = & - \sum_{i>j} \frac{q_i q_j}
{\left| \vec{\rho}_i - \vec{\rho}_j \right|} \, .
\end{eqnarray}
Clearly the Hamiltonian  (\ref{Hamphi})
describes  the one-dimensional motion of a particle
with the mass ${\cal M}$ and unit charge in the attractive field of
Coulomb center with the charge $- {\cal Q}_0$.
The corresponding equation of motion is given by
the considered previously
Eq.(\ref{alpha}) in which  the  constant $\alpha$
proves to be equal to
\begin{equation}\label{M/Q}
\alpha = \frac{{\cal Q}_0}{\cal M}.
\end{equation}
The interesting for us physical events take place
if there is enough Coulomb attraction in the
system. That is why we suppose that the
effective Coulomb charge ${\cal Q}_0$ is attractive,
${\cal Q}_0 > 0$, resulting in positive value of
$\alpha$.

Eqs.(\ref{M}),(\ref{Q}),(\ref{M/Q}) show
that arbitrary scaling of $\alpha$ can be compensated
for by the corresponding scaling of coordinates  $\rho_j$.
This fact agrees with  Eqs.(\ref{scaler}),(\ref{scalea}).

The scaling function $\phi(t)$ is defined by straightforward
integration of (\ref{alpha})
\begin{eqnarray} \label{En}
\frac{1}{2} \, {\cal M} \, \left( \frac{d \phi}{dt} \right)^2
- \frac{{\cal Q}_0}{\phi} = E ,
\end{eqnarray}
where $E$ is the system energy. Combined with the initial
condition $\phi(t_0) = 1$ this fixes the scaling function
unambiguously.

It is important to emphasize that
Eqs.(\ref{scale}), (\ref{QEC}) present
the idea of SC in an
invariant
form
independent
of
the
chosen coordinate frame.
To see this more
clearly
let us introduce
grand
vectors
in the $3N$ dimensional
configuration
space. The
grand
vector
${\bf r}(t) = (\vec r_1(t), \cdots \vec r_N(t))$
defines
the time-dependent coordinates,
the vector
$ \mbox{\boldmath $ \rho$ } =
(\vec \rho_1, \cdots \vec \rho_N)$
gives the initial coordinates
and
${\bf a} =(\vec a_1,\cdots \vec a_N)$
is
the vector
of
accelerations at the initial moment of time.
We employ
bold type to distinguish such a vector from the conventional
vector in space. 
Eqs.(\ref{scale}),(\ref{QEC}) allow the following presentation
\begin{eqnarray} \label{bfr}
{\bf r}(t) &=& \phi(t) \mbox{\boldmath $\rho$}, \\ \label{bfa}
\alpha \mbox{\boldmath $\rho$} &=& {\bf a}.
\end{eqnarray}
Obviously, these relations between $3N$-vectors
do not depend on a reference frame.  
This shows  that the  scaling coordinate $\phi(t)$
is described in an invariant way.

It has been presumed
by previous authors
that some coordinate which
describes fragmentation is to be singled out and
the potential extremum point is to be found for the fixed value
of this ``break-up coordinate''. The latter has been chosen in most cases
as the system hyperradius
\cite{Fano} \cite{FN} \cite{Peterkop} \cite{Rau}
defined as $R^2 = \sum_{i=1}^N m_i r_i^2$.
In the hyper-coordinate reference frame the potential energy
\begin{equation} \label{hyp}
V = \frac{C(\omega)}{R}
\end{equation}
is proportional to hypercharge $C(\omega)$ which
depends on a set of hyperangles $\omega = (\omega_1,\cdots \omega_{3N-7})$.
It is easy to verify that definition of the SC (\ref{bfa})
in the hyperspherical coordinates is reduced to
\begin{equation} \label{C'}
\frac{ \partial C(\omega)}{\partial \omega_i}=0,
\end{equation}
which shows that a SC is a saddle-point of the
hypercharge $C(\omega)$.
The function $\phi(t)$ in hyperspherical coordinates is proportional to
the hyperradius
$\phi(t) = R/R_0$, where $R_0$ is the initial value
of the hyperradius. The effective charge ${\cal Q}_0$ and the
effective mass ${\cal M}$ can be expressed in terms of $R_0$ and the
hypercharge ${\cal M} = R_0^2,~~{\cal Q}_0= -C_0/R_0$
where $C_0$ is the hypercharge evaluated for SC.
Description of the system in the hyper-coordinates
has a long tradition and list of achievements, see for the example
recent calculations of the three-electron
atom in hyperspherical coordinates \cite{Tolst}.
However, generally speaking, these coordinates do not
possess fundamental advantages over other coordinate frames
for the fragmentation problem.

Another well known  reference frame provide Jacoby coordinates
used in the approach  developed by Feagin \cite{F84}.
 For the multiparticle fragmentation
the choice of the ``break-up coordinate'' is not obvious
and some special procedure was developed for its construction
\cite{FF} \cite{Poel} \cite{Fantip}.
It becomes the more sophisticated problem the more complicated
the system is. 

In conclusion of this Section it should be stressed 
once more that our approach  provides an invariant 
definition for the idea of SC which is given
in (\ref{bfr}), (\ref{bfa}).

\section{SMALL DEVIATIONS FROM SCALING CONFIGURATION}
\label{small}
Assuming that the function $\phi(t)$ is defined as described in
the preceding section, we switch from $\vec{r}_j$ to the new
coordinates $\delta \vec{r}_j$
\begin{eqnarray}
\vec{r}_j = \phi(t) \, \vec{\rho}_j + \delta \vec{r}_j
\end{eqnarray}
which have an obvious meaning of deviations from the SC.
Presuming that these deviations are small, we write down
linearized classical Newtonian equations for $\delta \vec{r}_j(t)$ as
\begin{eqnarray}
m_i \, \frac{d^2 \delta \vec{r}_i}{dt^2} =
- \frac{1}{\phi(t)^3} \, \sum_{j=1}^N V_{ij} \, \delta \vec{r}_j,
\\ \label{V}
V_{ij} = \frac{ \partial^2}{\partial \vec{\rho}_i \,
\partial \vec{\rho}_j} \,
\sum_{m>n} \frac{q_m q_n}
{\left| \vec{\rho}_m - \vec{\rho}_n \right|} \, .
\end{eqnarray}
These equations of motion are generated by the time-dependent
Hamiltonian function
\begin{eqnarray}
\delta H = \frac{1}{2} \sum_{j=1}^N
\frac{\delta \vec{p}_j^{\, 2}}{m_j}+
\frac{1}{2\phi(t)^3} \, \sum_{i, \,j =1}^N V_{ij} \, \delta \vec{r}_i
\cdot \delta \vec{r}_j ,
\quad \quad \quad
\delta \vec{p}_j \equiv m_j \, \frac{\delta \vec{r}_j}{dt} .
\end{eqnarray}

It is convenient to introduce scaled deviations $\vec{\xi}_j$
and related momenta $\vec{\pi}_j$ as
\begin{eqnarray} \label{sca}
\vec{\xi}_j = \frac{1}{\phi(t)^{3/4}} \, \delta \vec{r}_j ,
\quad \quad \quad
\vec{\pi}_j = \phi(t)^{3/4} \, \delta \vec{p}_j ,
\end{eqnarray}
since this allows us to factor out the time dependence in the Hamiltonian:
\begin{eqnarray} \label{H1}
\delta H = \frac{1}{\phi(t)^{3/2}} \,
\left[ \frac{1}{2} \sum_{j=1}^N \frac{\vec{\pi}_j^{\, 2}}{m_j} +
\frac{1}{2} \, \sum_{ij} V_{ij} \, \vec{\xi}_i \, \vec{\xi}_j
- \frac{3}{8} \sqrt{\phi} \, \frac{d \phi}{dt} \,
\sum_{j=1}^N \left( \vec{\xi}_j \cdot \vec{\pi}_j +
\vec{\pi}_j \cdot \vec{\xi}_j \right)
\right] .
\end{eqnarray}
The derivation of this formula could be traced via
a quantum mechanical analogue of the problem (which
for many readers nowadays is more convenient
than  the pure classical consideration).
In quantum mechanics the transformation rules for momenta
and the Hamiltonian follow respectively from the formulae for
the partial derivatives
\begin{eqnarray}
\frac{\partial}{\partial \vec{r}_j} = \phi(t)^{-3/4}
\frac{\partial}{\partial \vec{\xi}_j} ,
\quad \quad \quad
\left( \frac{\partial}{\partial t} \right)_{\delta \vec{r}_j} =
\left( \frac{\partial}{\partial t} \right)_{\vec{\xi}_j} +
\frac{3}{4 \phi(t)^{1/4}} \, \frac{d \phi}{dt} \,
\sum_{j=1}^N \vec{\xi}_j \cdot \frac{\partial}{\partial \delta \vec{r}_j} .
\end{eqnarray}
In (\ref{H1}) we use a symmetrized representation which should be employed
in the quantum version of the formulae (the latter also implies a
corresponding gauge transformation for the wave function).

 From (\ref{En}) one obtains
\begin{eqnarray}\label{ene}
\sqrt{\phi} \: \frac{d \phi}{dt} = \sqrt{\frac{2 (E \phi(t) +
{\cal Q}_0)} {{\cal M}}}
\end{eqnarray}
which becomes time-independent for $E=0$. In this case the
time-dependence is {\it exactly}\/ factored out in the
Hamiltonian (\ref{H1})  justifying the choice of the scaling
(\ref{sca}). This implies that the original
{\it non-stationary}\/ problem becomes {\it stationary}\/
provided one replaces time $t$ by an effective time $\tau$.
A relation between $t$ and $\tau$ in differential form is
\begin{eqnarray} \label{tau}
d \tau = \phi(t)^{-3/2} \, dt .
\end{eqnarray}
For some applications it is necessary to keep the energy dependence
of the trajectory. For these cases a convenient technique
has been developed recently by Kuchiev \cite{Kuch}.
We have applied it to the case considered and verified that
it results in the same threshold indexes as the
ones obtained below by the stationary approach.

The Hamiltonian describing propagation in the effective time (\ref{tau})
reads
\begin{eqnarray} \label{Htau}
&&\delta H_{\tau} =
\frac{1}{2} \sum_{j=1}^N \frac{\vec{\pi}_j^{\, 2}}{m_j} +
\frac{1}{2} \, \sum_{ij} V_{ij} \, \vec{\xi}_i \, \vec{\xi}_j +
\frac{a}{2} \sum_{j=1}^N \left( \vec{\xi}_j \cdot \vec{\pi}_j +
\vec{\pi}_j \cdot \vec{\xi}_j \right),
\\
\label{a}
&&a = - \frac{3}{4} \, \sqrt{\phi} \, \frac{d \phi}{dt} .
\end{eqnarray}

The Hamiltonian $\delta H_\tau$ (\ref{Htau}) is quadratic in
coordinates and momenta thus describing  a set of harmonic
oscillators or inverted oscillators.
This shows that our goal is to describe the behavior of the system
in terms of these oscillators and inverted oscillators.
Before proceeding we modify our notation. The set of components
of the displacements vectors $\delta \vec{r}_j$ ($j=1, 2, \ldots N$)
comprise $3N$-dimensional grand vector ${\bf \delta r}$.
In this formulation, for instance,
$V_{ij}$ corresponds to grand $3N \times 3N$ square matrix
denoted below as ${\bf V}$.
We introduce also $3N \times 3N$ unit matrix
${\bf I}$ and the diagonal matrix ${\bf K}$ of the same size with
diagonal elements corresponding to inverse mass $1/m_j$ of each particle.

This notation takes into account an obvious fact that
the total number of all modes  coincides with
the number of degrees of freedom in the system
($k = 1, \; 2, \; \ldots 3 N$).
There are however seven particular degrees of freedom:
translations, rotations and the scaling transformation.
They do not change  the shape of a SC and
do not describe a deviation  from a SC.
These degrees of freedom may be called
the collective modes.
They  obviously  should be considered separately
from the oscillating modes which describe deviations from
the SC. In order to distinguish the collective modes one can
use the following interesting property.
All collective degrees of freedom
are described by the eigenvectors
of the grand matrix ${\bf K V}$ with particular
eigenvalues. Firstly, the three modes which
correspond to the system translations in space
have obviously zero eigenvalues.
Secondly,
the modes corresponding to rotations of the system in space
have eigenvalues equal to ${\cal Q}_0/{\cal M}$,
as shown in Appendix.
There are three such modes in general case, while
for a linear SC there are only two modes.
Thirdly, the mode
corresponding to the scaling transformation
Eqs.(\ref{scaler}),(\ref{scalea}) has an eigenvalue
$-2{\cal Q}_0/{\cal M}$, as also shown in Appendix.
Using these eigenvalues one can separate the
collective modes either from the very beginning, or at the
end of calculations.

There is another useful for applications way to separate the
collective modes.
For translations and rotations the
separation can  be fulfilled by conventional methods choosing
appropriately the coordinates,
as is demonstrated in a number of examples below.
Separation of the  scaling  mode
can be achieved
with the help of the operator of projection on this mode ${\bf P}$
and the complementary projection operator ${\bf Q} = {\bf I} - {\bf P}$.
The operator ${\bf P}$ is readily constructed from the unit vectors
$\vec{n}_j = \vec{\rho}_j/\rho_j$ which define the shape of SC:
\begin{eqnarray}
P_{ij} = \vec{n}_i \cdot \vec{n}_j .
\end{eqnarray}
Thus all seven collective  modes can be  easily
identifies and separated using any of the two
techniques described above.

Some modes in the  $3N-7$ subspace orthogonal to the collective modes
are stable and describe
small oscillations around SC; the related
oscillating frequencies $\omega_k$
are real. The object of our major interest is
unstable modes with imaginary
oscillating frequencies.
It is shown below that  unstable modes exist for any SC.
It is convenient to introduce
for unstable modes a
parameter $\alpha_k = i \omega_k$ (${\rm Re} \, \alpha_k >0$).
In order to find the oscillating
frequencies one can  presume a harmonic time-dependence
of the coordinates $\xi$ and momenta $\pi$
\begin{eqnarray}
{\mbox {\boldmath $\xi$} }
= \exp (i \omega t)  \, {\bf \Xi} ,
\quad \quad \quad
{\mbox {\boldmath $\pi$} }
 = \exp (i \omega t)  \, {\bf \Pi} ,
\end{eqnarray}
where ${\bf \Xi}$ and ${\bf \Pi}$ are time-independent
grand vectors. The Hamiltonian equations of motion give
\begin{eqnarray}
i \omega \, {\bf \Xi} = {\bf K}  \, {\bf \Pi}
+ a \, {\bf \Xi} ,
\quad \quad \quad
i \, {\bf \Pi} = - {\bf V}  \, {\bf \Xi}
- a \, {\bf \Pi} ,
\end{eqnarray}
where $a$ (\ref{a}) is a scalar coefficient.
The latter equation could be written also as
\begin{eqnarray}
i \omega \left( \begin{array}{c}
{\bf \Xi} \\ {\bf \Pi}
\end{array} \right)
=
\left( \begin{array}{c c}
a  &   {\bf K}  \\
-  {\bf V}  & - a  \end{array} \right)
\left( \begin{array}{c}
{\bf \Xi} \\ {\bf \Pi}
\end{array} \right) .
\end{eqnarray}
Excluding the grand vector ${\bf \Pi}$ one comes to the
eigenvalue problem for the square of frequency $\omega^2$
\begin{eqnarray} \label{EV}
(\omega^2 + a^2)  \, {\bf \Xi} =
{\bf K} \, {\bf V}  \, {\bf \Xi} \, ,
\end{eqnarray}
or, in the symmetrized form
\begin{eqnarray}
(\omega^2 + a^2) \, \tilde{{\bf \Xi}} =
{\bf K}^{1/2}  \, {\bf V}  \, {\bf K}^{1/2} \,
\tilde{{\bf \Xi}} \, ,
\quad \quad \quad
{\bf K}^{1/2} \, \tilde{{\bf \Xi}} =   {\bf \Xi} \, .
\end{eqnarray}
Denoting a set of  eigenvalues of the matrix
${\bf K}  \, {\bf V} $
as $v_k,~k=1,2,\cdots,3N$, we obtain
\begin{eqnarray}
\omega_k^2 = v_k - a^2 , \\
\alpha_k = \sqrt{a^2 - v_k} .
\end{eqnarray}
This formula shows how
the oscillation frequencies depend on the eigenvalues
of the matrix ${\bf K}  \, {\bf V} $.

Let us verify now that a SC is always unstable.
With this purpose let us show that the matrix ${\bf K \,V}$
always possesses negative eigenvalues which
describe instability.
Consider
the trace of the grand matrix ${\bf V}$
\begin{eqnarray}\label{trgr}
{\rm Tr} \, {\bf V} = \sum_j \, \frac{ \partial^2}
{\partial \vec{\rho}_j \, \partial \vec{\rho}_j} \,
\sum_{m>n} \frac{q_m q_n}
{\left| \vec{\rho}_m - \vec{\rho}_n \right|} \, =
\sum_j \, \bigtriangleup_{\vec{\rho}_j} \,
\sum_{m>n} \frac{q_m q_n}
{\left| \vec{\rho}_m - \vec{\rho}_n \right|} = 0,
\end{eqnarray}
which vanishes
since the Coulomb potential satisfies the Laplace equation
\begin{eqnarray}\nonumber
\bigtriangleup_{\vec{\rho}_j} \, \frac{1}
{\left| \vec{\rho}_j - \vec{\rho}_n \right|} = 0
\quad \quad \quad
\vec{\rho}_j \neq \vec{\rho}_n .
\end{eqnarray}
It is easy to see also that (\ref{trgr}) results in
${\rm Tr} \, ({\bf K \, V}) = 0$ which means that
\begin{eqnarray}\label{trace}
{\rm Tr}\,( {\bf K \, V}) = \sum_k v_k = 0 .
\end{eqnarray}
We see that the spectrum of the matrix ${\bf K \, V}$ always contains
both  positive and negative eigenvalues.
This fact in itself is not sufficient to make a statement
about instability because
the trace (\ref{trace}) includes contribution
from collective modes which do not change a shape
of the SC. However, it is easy to exclude collective modes.
 Remember that the eigenvalues
corresponding to translations are zero, rotations give eigenvalues
${\cal Q}/{\cal M}$, while the scaling transformation
provides the eigenvalue $-2 {\cal Q}/{\cal M}$, see Appendix.
The sum of eigenvalues of collective modes is
\begin{equation}\label{trcol}
\sum_{\rm collective~modes}v_k = \left\{
\begin{array}{rr} \cal Q / \cal M~~~~ {\rm in~ general~ case}\\
0~~~~~~{\rm for~ linear~ SC}.
\end{array} \right.
\end{equation}
Subtracting this result from (\ref{trace}) we find
the  trace of the matrix ${\bf K \, V}$  in the subspace orthogonal
to the collective modes
\begin{equation}\label{orth}
{\rm Tr}\, ({\bf K\,V})_{\rm orth} = \sum_{\rm orthogonal} v_k =
\left\{\begin{array}{rr}
- \cal Q / \cal M~~~~ {\rm in~ general~ case}\\
0~~~~~~{\rm for~ linear~ SC}.
\end{array} \right.
\end{equation}
Since this trace is non-positive, we
conclude that the matrix ${\bf K \, V}$ inevitably   possesses
negative eigenvalues which describe deviations from the SC.
This shows that any SC is unstable.
This property is closely related
to the fact that harmonic functions, i.e. those which satisfy
the Laplace equation, cannot have maxima or minima.
Remember also the Earnshow
theorem well known in electrostatics: stable equilibrium is
impossible for systems where the Coulomb forces are operative.
Although SCs describe expanding non-static
configurations, conclusion about inevitable 
instability remains valid in this case as well.
This fact can be interpreted 
as a dynamic analogue of the Earnshow theorem.

\section{QUANTIZATION OF DEVIATIONS FROM SCALING
CONFIGURATION AND THRESHOLD INDICES}
\label{quant}

Previous section reduces description of small deviations from SC
to the set of coupled harmonic oscillators which could be
quantized straightforwardly. This procedure
provides the 'energy' levels
\begin{eqnarray}
\epsilon_{k n_k} = \omega_k \left( n_k + \frac{1}{2} \right) .
\end{eqnarray}
Here the first subscript $k=1,2,\cdots,3N-7$
indicates the mode, and $n_k= 0,1, \cdots$ shows a number
of quanta in this mode.
For a given set of the quantum numbers $\{n_k\}$ the system wave
function is given by
\begin{eqnarray} \label{Psi}
\Psi_{ \{ n_k \}} \sim
\exp \left( - \sum_k i \int_{\tau_0}^\tau
\epsilon_{k n_k} \, d \tau \right) =
\exp \left( - \sum_k i \int_{t_0}^t
\frac{\epsilon_{k n_k}}{\phi(t)^{3/2}} \, d t \right) ,
\end{eqnarray}
where we omit the common time-dependent phase factor.
The wave function is prepared at some initial moment $t_0$ by
preceding strong interaction of all fragments.
In the Wannier-type approach it is presumed that these processes
depend smoothly on the energy $E$. Hence they do not influence the form
of threshold law and thus could be effectively excluded from consideration;
it is sufficient to consider only $t > t_0$ domain.

For unstable modes
the 'energies' $\epsilon_{k n_k}$ are complex-valued which
leads to the loss of probability in the expanding SC.
This should be interpreted \cite{KOJETP} \cite{KOharm}
as sliding from the potential saddle
in multidimensional configuration space that eventually
leads to formation of bound states of two (or more) fragments.
Such an outcome implies that the
related part of probability is lost for the process of
complete system fragmentation which is an object of our study.
Cross section of the latter is proportional to the
{\it survival probability}
\begin{eqnarray} \label{P}
P_{ \{ n_k \}} \equiv \left|
\Psi_{ \{ n_k \}} |_{t \rightarrow \infty} \right|^2 =
\exp \left(- \sqrt{ 2 {\cal M}} \: \sum_k \int_{\phi(t_0)}^\infty
\frac{\alpha_{k n_k}}{\phi \sqrt{E \phi + {\cal Q}_0}} \, d \phi
\left( n_k + \frac{1}{2} \right)
\right) ,
\end{eqnarray}
where summation over $k$ runs over all unstable modes.
Note that the original quantum problem is stationary. The time
$t$ in (\ref{P}) plays a role of an effective variable which
describes  scaling of the system in accordance with (\ref{En}).
Small deviations from SC are described quantum mechanically.
Our treatment generalizes to multimode
case the scheme developed by Kazansky and Ostrovsky \cite{KO92}
for the two-electron escape (see also Ref.\cite{KOZPhys}; some
of ideas used were elaborated also by Watanabe \cite{Wat91}).
Note that the cited paper \cite{KO92}
provides also a description of deviations
from pure power threshold law, but we do not pursue this point here.

For our objectives it is sufficient to note that (\ref{P})
has a form of a product of contributions coming from each individual
mode, hence the threshold law of interest is
%
%
\begin{eqnarray} \label{LAW}
\sigma \sim P_{ \{ n_k \}} \sim E^ \mu,
\\
\label{ind}
\mu \equiv { \sum_k \mu_{k n_k}} .
\end{eqnarray}
%
The {\it partial threshold indices}\/ $\mu_{k n_k}$ stem from
the 'eigenfrequencies' of unstable modes being related to
the {\it negative}\/ eigenvalues $v_k < 0$ of the ${\bf K \, V}$ matrix
\begin{eqnarray} \label{partind}
\mu_{k n_k}= 2 \left[\sqrt{ - \frac{{\cal M}}{2 {\cal Q}_0} \, v_k
+ \frac{9}{16}} - \frac{1}{4} \right]
\left( n_k + \frac{1}{2} \right) .
\end{eqnarray}
Small positive values of $v_k$ formally could also lead to
real $\mu_{k n_k}$, but have to be discarded.
Obviously, if some (imaginary) 'eigenfrequencies' are $N_k$-fold
degenerate, the related contributions appear $N_k$ times
in the sum (\ref{ind}).
In principle the wave function is a superposition of terms
corresponding to various sets of quantum numbers $\{ n_k \}$,
since all of them are populated by the processes in the
inner interaction domain. Clearly, the threshold law is defined
by the least possible values of $n_k$ \cite{angmod}
which are equal to zero
unless the symmetry considerations forbid this
choice, as exemplified in the next paragraph.
If the initial SC is scaled by the factor $\lambda$, see
Eqs.(\ref{scaler}), (\ref{scalea}), then
$ {\bf K \, V} \sim \lambda^{-3}$, ${\cal M} \sim \lambda^2$,
${\cal Q}_0 \sim \lambda^{-1}$, but the threshold indices $\mu_{k n_k}$,
as anticipated, remain scale-independent. Note also that the threshold
index is invariant under simultaneous scaling of all charges
or all masses in the system.

In the original Wannier problem two electrons escape from
infinitely heavy atomic core with the charge $Z$. The configuration
found by Wannier \cite{Wannier}
gives the simplest example of SC in which
the electrons reside at equal distances $\rho$ and in opposite
directions from the core.
The motion is unstable with respect to the
stretching mode which is separated from the (stable) bending mode.
Thus it is sufficient for our purposes to consider
motion of electrons along the line passing through the core.
This motion is described by two coordinates and the matrix
${\bf V}$ takes the form
\begin{eqnarray}
{\bf V} = \frac{1}{\rho^3} \left( \begin{array}{c c}
-2 Z + \frac{1}{4} & - \frac{1}{4} \\
- \frac{1}{4} & -2Z + \frac{1}{4}
\end{array} \right)
\nonumber
\end{eqnarray}
and
${\cal Q}_0 = \left(2Z - \frac{1}{2}\right)\rho^{-1}$,
${\cal M} = 2 \rho^2$ (we use atomic system of units, ${\bf K=I}$).
The eigenvalues of ${\bf V}$ are $v_1 = -2Z/\rho^3$ and
$v_2 = \left( -2Z + \frac{1}{2} \right) /\rho^3$.
The eigenvalue $v_2$ is seen to coincide with $- 2 {\cal Q}/{\cal M}$.
Hence it corresponds to SC expansion (see Appendix) and should be
discarded. The eigenvalue $v_1$ upon substitution into (\ref{partind})
reproduces the well known result
\begin{eqnarray}
\mu_{1 n_1}= \frac{1}{2} \left[\sqrt{\frac{100Z-9}{4Z-1}} - 1 \right]
\left( n_1 + \frac{1}{2} \right) .
\end{eqnarray}
The choice $n_1=0$ provides the famous Wannier law valid for
$^1S$ symmetry of the final two-electron continuum state, whereas
$n_1 =1$ corresponds to the threshold law for $^3S^e$ (and $^3P^e$)
symmetry \cite{triplet}.

Feagin and Filipczyk \cite{FF} and Poelstra {\it el al}\/ \cite{Poel}
put forward another formula for the threshold index in the multimode case.
According to it the Wannier index is $(N-2)$ times larger than (\ref{ind}).
The factor $(N-2)$ is described as a ``phase space factor for $(N-1)$
outgoing particles'' being justified by the reference to the earlier
paper by Feagin \cite{F84}. We were unable to find the derivation of
such a factor in the cited paper; anyway it deals only with the
conventional $N=3$ case where the factor $(N-2)$ is insignificant.
Our treatment provides purely dynamic approximation for
the wave function and does not leave any room for the statistical
arguments. The other aspects of relation between dynamic and
statistical threshold laws are discussed in Section VI.

\section{PARTICULAR SYSTEMS}
\label{parti}
In practical applications of our scheme the less obvious part
corresponds to finding SCs. Numerical solution of the set of
non-linear equations (\ref{QEC}) could be cumbersome and implies
reasonable initial guess. The question whether all the
solutions are found is even more difficult. In reality
one has to appeal to intuitive reasoning and to limit
search to some symmetrical configuration. This allows one to
effectively reduce the number of equations (\ref{QEC})
to be considered.
Since the initial step of finding SC in most cases could not be
done in closed form,
we do not pursue the goal of obtaining analytical formulae,
but resort to numerical calculations which are performed using
the {\it Mathematica}\/ \cite{Mat} program.
We find it easier to avoid preliminary
separation of rotational and translational coordinates,
since they could be easily distinguished in the eigensystem
of the complete matrix ${\bf K}  \, {\bf V} $.
Moreover, the known eigenvalues of this matrix corresponding to
rotations (see Appendix) provide a good test for consistency
of calculations.

The systems practically accessible nowadays in atomic physics
are not very diverse, consisting of several electrons and positrons
in the field of heavy (positively charged) atomic core.
Since three-particle systems (such as
${\rm A}^{+Z} + 2e^-$ or ${\rm A}^{+Z} + e^- + e^+$)
are already studied in great detail
\cite{Wannier} \cite{Peterkop} \cite{Rau} \cite{Klarpos}
\cite{KlarZPhys} \cite{Gru82} \cite{Grumass}
(see also references in the Introduction), we start from
the four-particle systems.
We do not impose any symmetry constraints on the system state
thus presuming that $n_k =0$ for all modes contributing $\mu$ (\ref{ind}).

\subsection{Three-electron escape from the charged core}

The system ${\rm A}^{+Z} + 3e$ was thoroughly investigated by Klar
and Schlecht \cite{KS} and Gruji\'{c} \cite{Gru3e}.
They considered a configuration of electrons
forming  an equilateral triangle with infinitely
massive core in the center, which is obviously a SC.
The out-of-plane motion is separated.
It corresponds to stable modes and does not affect the threshold law.
The in-plane motion is described by six coordinates of electrons,
or by four 'oscillatory' modes plus uniform expansion of SC and its
rotation. The eigenfrequencies obtained by us, as
well as in the cited papers,
are pairwise degenerate due to
SC symmetry. One pair corresponds to stable motion and the other
pair to unstable motion. The latter pair produces two equal terms
in the sum (\ref{ind}). Klar and Schlecht \cite{KS} and Gruji\'{c}
\cite{Gru3e} succeeded
in deriving analytical expressions for the Wannier index \cite{f1}.
In this paper we do not pursue analytical formulations but
check that our numerical results coincide with those cited by
Gruji\'{c}, namely $\mu = 2.82624$ for $Z=1$,
$\mu = 2.27043$ for $Z=2$, $\mu = 2.16196$ for $Z=3$, etc.
The experiment for electron impact double ionization of atoms ($Z=3$)
seem to agree with the threshold law \cite{exp3e}.

The two pairs of modes discussed above are already well known.
Combined with rotation and scaling expansion they
represent a complete set of six in-plane coordinates.
Since the number of modes is a physical parameter
which is independent on the theoretical technique used, we do not see
any possibility to obtain some additional unstable
modes which would lead to another Wannier index and thus to the
complementary threshold law as announced by Feagin and Filipczyk
\cite{FF} (in fact our conclusion could be drawn from the
paper by Gruji\'{c} \cite{Gru3e} who used the plain Cartesian
coordinates whereas less transparent treatment by Klar and
Schlecht \cite{KS} is based on hyperspherical coordinates).
Since no details of analysis by Feagin and Filipczyk \cite{FF}
were ever published, more detailed discussion of this issue is not possible.

\subsection{$2 e^- + e^+$ escape from the charged core}

The plausible symmetric SCs for the system
${\rm A}^{+Z} + 2e^- + e^+$ were considered by
Poelstra {\it et al}\/ \cite{Poel}
(note that the calculations in this paper
were carried out only for $Z=1$). They comprise two different
linear arrangements and one plain configuration
\cite{GruFizCom}.
All these configurations belong to SC and therefore
can be easily handled by the developed above technique.
We consider below these SCs
successively.

\subsubsection{Linear configuration $L_a$}

Let the frame origin is placed
into infinitely massive core having the charge $Z$.
The coordinates of two electrons and
positron are respectively $x_1$, $x_2$, $x_3$; all of them are positive.
It is convenient to introduce two dimensionless parameters
$x = r_1/r_3$ and $y = r_2/r_3$ $(0 < x < 1 < y)$
which have to satisfy the system
of equations obtained from (\ref{QEC})
\begin{eqnarray}
\frac{m_3}{m_1} \, \frac{Z/x^2 - 1/(1-x)^2 + 1/(y-x)^2}
{-Z + 1/(1-x)^2 - 1/(1-y)^2} = x ,
\nonumber \\
\frac{m_3}{m_2} \, \frac{Z/y^2 + 1/(1-y)^2 - 1/(y-x)^2}
{-Z + 1/(1-x)^2 - 1/(1-y)^2} = y
\end{eqnarray}
(equations are presented  for  more general case when all
light particles have different masses $m_i$, while  the core remains
infinitely heavy).

\subsubsection{Linear configuration $L_b$}

A distinction from the previous case is that the  coordinate
of one of the electrons is negative ($x_2 <0$). The system of equations
defining SC is somewhat different $(y < 0 < x < 1)$:
\begin{eqnarray}
\frac{m_3}{m_1} \, \frac{Z/x^2 - 1/(1-x)^2 - 1/(y-x)^2}
{-Z + 1/(1-x)^2 + 1/(1-y)^2} = x ,
\nonumber \\
\frac{m_3}{m_2} \, \frac{-Z/y^2 - 1/(1-y)^2 + 1/(y-x)^2}
{-Z + 1/(1-x)^2 + 1/(1-y)^2} = y .
\end{eqnarray}
For both linear configurations the bending modes are stable.
There are two stretching modes for each configuration, both
being unstable. The results of our calculations are
summarized  in table 1. For $Z=1$  parameters $x$, $y$
and partial threshold indices $\mu_1$ and
$\mu_2$ coincide with those obtained by Poelstra {\it et al} \cite{Poel};
our threshold indices $\mu$ are less by a factor of 2, as discussed
at the end of Section \ref{quant}. Notice a  non-trivial
behaviour of the parameters with $Z$: for instance, in SC
$L_a$, $x$ and $\mu_1$ increase
with $Z$, whereas $y$ and $\mu_2$ decrease. The threshold index
$\mu$ increases with $Z$ which is opposite to the well known
behaviour for the simplest system (${\rm A}^Z + 2e$) and
for $3e$ escape where $\mu$ diminishes as $Z$ grows
(see more discussion in the Section \ref{disc}).

\subsubsection{Plane configuration $P$}

The symmetric plane configuration is conveniently
characterized by two angles $\alpha$ and $\beta$:
$\alpha$ is an angle between two lines which join
the ion with the positron and one of the electrons,
while
$\beta$ is an angle between two  lines which join  
the positron with the ion and with one of the electrons.
In the considered plane configuration the two electrons
are located symmetrically, which  means that
their locations mirrow each other under
 reflection in an axes which joins the ion and the positron.
This makes  the angles
$\alpha$ and $\beta$ be identical for both electrons.
From (\ref{QEC})
we deduce the system of equations
\begin{eqnarray} \label{2ep}
m_+ \, \sin^3 \gamma \left( Z \, \sin^2 \alpha  -
\frac{1}{4} \sin \alpha +
\sin^2 \beta \, \cos \gamma \right)  =
\nonumber \\ =
m_- \sin^3 \beta \left( 2 \cos \beta \, \sin^2 \gamma -
Z \sin^2 \alpha \right) ,
\nonumber \\
\sin^2 \beta \, \sin \gamma = \frac{1}{4} \, \cos \alpha
\quad \quad \quad (\gamma = \pi - \alpha - \beta) ,
\end{eqnarray}
where the masses of light particles with negative ($m_-$) and
positive ($m_+$) charges generally could be different.
The results of calculations are presented in table 1. For $Z=1$
the angles $\alpha$ and $\beta$ coincide with these extracted from
the paper by Poelstra {\it et al}\/ \cite{Poel}.
However, the difference between the threshold indices
is drastic. Poelstra {\it et al}\/ had found a single
unstable mode which corresponds to our partial Wannier index $\mu_2$.
Our calculations give two unstable modes, similarly to
the case of $3e$ escape (in the latter case the modes were
degenerate due to a symmetry which is absent for the system under
consideration).
The reason of this disagreement remains unclear.

The plane SC $P$ governs the threshold behaviour, although it
provides the threshold index $\mu$ only slightly less than the
linear configuration $L_b$.


\subsection{Four-electron escape from the charged core}

Basing on the symmetry considerations we analyze three configurations:
linear, plane and 3D SC. It could be shown rigorously that
for symmetric linear arrangement SC does not exist for all values
of $Z$, i.e. Eqs.(\ref{QEC})  have no solution.


\subsubsection{Plane configuration $P$}

In the plane configuration the electrons are located in the
apexes of a square; the core lies in its center.
The out-of-plane motion is separated and
corresponds to stable modes. For in-plane motion
in general case we  find one non-degenerate
and one doubly-degenerate unstable modes (table 2).
For the particular case
$Z=1$ an additional non-degenerate mode becomes unstable.

\subsubsection{{\rm 3D} configuration $V$}

SC describes the electrons located at the apexes of
tetrahedron. We find a single triply-degenerate mode (table 2).
Interestingly, the threshold index $\mu$ proves to be quite close
for plane and 3D configurations, although 3D SC provides
somewhat lower value of $\mu$ and thus governs the threshold
behaviour. As $Z$ increases, the relative importance of
electron-electron interaction decreases and $\mu$ approaches
the value $\mu = 3$ which corresponds
to non-interacting electrons.

The smallest practically attainable
value of the charge seems to be $Z=2$. It could be realized via triple
ionization of negative ion by electron impact.
However, theoretically the case $Z=1$ proves be very
interesting due to unusual properties.
In this case the threshold index becomes much larger than in other
cases, particularly for the plane SC.
This is due to a  small value of the
'charge' ${\cal Q}_0$ in this case. Another interesting feature is an
appearance of an additional unstable mode in the plane SC.
An analysis of the eigenvector ${\bf \Xi}$ shows that it corresponds
to the out-of-plane motion. Namely, a pair of electrons lying on a diagonal
of the square shifts upwards, whereas another pair shifts downwards.

The tetrahedric configuration was considered earlier
by Gruji\'{c} \cite{Gru4e} who obtained approximate analytical
expressions for the threshold indexes. The partial threshold indexes
obtained by him reveals only an approximate degeneracy.
The numerical results for $\mu$
are in reasonable agreement with our data.

\subsection{$3 e^- + e^+$ escape from the charged core}

We failed to find a symmetrical plane SC for this system.

\subsubsection{Linear configuration $L$}

Linear SC  corresponds to alternating positive and negative
charges. Let us locate the origin at the heavy ion
and call by $x_1 >0,~~ x_2 < 0, ~~x_3> x_1 >0$
locations of three electrons, and by
$x_4,~~x_1< x_4 <x_3 $  location of the positron.
Then the considered configuration
can be characterized by three parameters:
$x=x_1/x_4$, $y=x_2/x_4$, $z=x_3/x_4$ $((y < 0 < x <1 < z$).
They have to satisfy a set of equations which follow from
(\ref{QEC})
\begin{eqnarray}
- \frac{Z}{x^2} + \frac{1}{(x-y)^2} + \frac{1}{(1-x)^2}
- \frac{1}{(z-x)^2} = x \left[ Z - \frac{1}{(1-x)^2}
+\frac{1}{(z-1)^2} - \frac{1}{(1-y)^2} \right] ,
\nonumber \\
\frac{Z}{y^2} - \frac{1}{(x-y)^2} + \frac{1}{(1-y)^2} - \frac{1}{(z-y)^2}
= y \left[ Z - \frac{1}{(1-x)^2}
+\frac{1}{(z-1)^2} - \frac{1}{(1-y)^2} \right] ,
\nonumber \\
- \frac{Z}{z^2} + \frac{1}{(z-x)^2} + \frac{1}{(z-y)^2}
- \frac{1}{(z-1)^2} =
z \left[ Z - \frac{1}{(1-x)^2}
+\frac{1}{(z-1)^2} - \frac{1}{(1-y)^2} \right] .
\end{eqnarray}
The parameters of SC and the Wannier indices are shown in table 3.

\subsubsection{{\rm 3D} configuration $V$}

The symmetrical 3D configuration arises when the three
electrons form the
equilateral triangle while the ion and the positron are located
up and down the plane of the triangle
on the  perpendicular to the triangle plane  which
crosses its center. 
 This configuration  is characterized
by two angles $\alpha$ and $\beta$ defined 
similar to the case considered in 
Section B3. Namely,
$\alpha$ is an angle between two lines which join
the ion with the positron and with one of the electrons, while
$\beta$ is an angle between two  lines which join  
the positron with the ion and with one of the electrons.
These angle are difined 
by equations similar to
(\ref{2ep}):
\begin{eqnarray}
m_+ \,  \sin^3 \gamma \left( Z \sin^2 \alpha -
\frac{1}{\sqrt{3}} \sin \alpha + \sin^2 \beta \, \cos \gamma \right) =
\nonumber \\ =
m_- \, \sin^3 \beta \left( 3 \cos \beta \, \sin^2 \gamma -
Z \sin^2 \alpha \right) ,
\nonumber \\
\sin^2 \beta \, \sin \gamma = \frac{1}{\sqrt{3}} \cos \alpha
\quad \quad \quad (\gamma = \pi - \alpha - \beta) .
\end{eqnarray}
We have found two doubly-degenerate and one non-degenerate
unstable mode as shown in table 3. The threshold law is
governed by 3D SC $V$. Note that the threshold index grows
with $Z$.

\subsection{Five-electron escape from the charged core}

\subsubsection{Plane configuration $P$}

In the plane SC the electrons are located in the
apexes of a equilateral pentagon; the core lies in the same plane.
In the in-plane motion we have found two doubly degenerate
unstable modes (table 4). For $Z=2$ an additional pair of
unstable modes appears.

\subsubsection{{\rm 3D} configuration $V$}

Here three electrons lie in the apexes of equilateral triangle
with the core in its center. On the perpendicular to this plane,
above the plane and below it, another pair of electrons
is located symmetrically.
The SC can be characterized by the angle $\alpha$ between the
line which joins out-of-plane electron with the core and the
line which joins it with in-plane electron. The angle is defined
by the equation
\begin{eqnarray} \label{5e3D}
\frac{1}{\sqrt{3}} + 2 \sin^3 \alpha - Z =
\tan \alpha \left( 3 \sin^2 \alpha \, \cos \alpha +
\frac{1}{4} \tan^2 \alpha - Z \tan^2 \alpha \right) .
\end{eqnarray}
Quite unexpectedly, $\alpha$ proves to be very close to 45$^\circ$,
exhibiting weak dependence on the core charge $Z$ (table 4). This means
that in-plane and out-of-plane electrons are located at almost
the same distance from the core. The 3D SC generates somewhat
lower values of $\mu$ than the plane SC thus governing the threshold
behaviour. However, the difference is quite small. This feature is
common to that found above for the four-electron case.

3D configuration for five-electron system was considered
previously by Dmitrievi\'{c} {\it et al} \cite{Gru5e}.
However, the equation derived for SC angle $\alpha$
differs from (\ref{5e3D}).

\subsection{Fragmentation in two pairs of identical particles
with opposite charges}

In this subsection we consider fragmentation into the final state
$2 \, X^{+Z}_m + 2e$, where $X^{+Z}_m$ is a positively charged
particle with charge $Z$ and mass $m$ (all results below hold if
the electrons are replaced by any other charged particles;
then $Z$ and $m$ have the meaning of ratio of charges and masses
respectively). In the applications considered above the zero
eigenvalues of the matrix ${\bf V}$ do not emerge due to
the presence of infinitely massive core. In the
($2 \, X^{+Z}_m + 2e$) system such modes are present.
Another distinction is that for equal masses of leptons
in previous applications we have  always had ${\bf K} = {\bf I}$
and  $v_k$ have been the eigenvalues of the
$ {\bf V} $ matrix.
Now we have to diagonalize the complete matrix
${\bf K}  \, {\bf V} $.
Both these features do not create substantial difficulties.

    From the symmetry considerations it is clear
that a shape of the SC is a rhombus with the angle $2 \alpha$ at
the apexes where the particles $X^Z_m$ are situated.
The single SC parameter $\alpha$ is defined by the equation
\begin{eqnarray}
8 Z - \frac{Z}{\cos^3 \alpha} =
m \left( 8 Z - \frac{1}{\sin^3 \alpha} \right).
\end{eqnarray}
which follows from (\ref{QEC}).
Several examples are shown in  table 5. The simplest practical
realization is  the complete fragmentation of H$_2$ molecule by
photons where $\alpha$ is close to $30^\circ$ in agreement
with Feagin and Filipczyk \cite{FF} and the threshold index
proves to be huge.
Apparently this threshold behaviour could not be observed
in experiments \cite{Kos}. Another feasible realization with
moderate Wannier index is ionization of negative positronium ion
by positron impact ($Z=1$, $m$ =1).

We fail to find the linear configuration discussed by Stevens and
Feagin \cite{FBul}.

\section{DISCUSSION AND CONCLUSION}
\label{disc}

This  paper  formulates the idea of the SC.
Defined by (\ref{scale}), the SC is shown to arise when
a nonlinear set of Eqs.(\ref{QEC}) is satisfied.
Propagation of the system in the vicinity of SC
configuration governs the threshold law which
is found in
Eqs.(\ref{LAW}),(\ref{ind}) and (\ref{partind}).
These results permit direct
practical calculations of the threshold index $\mu$
for any system.

In many cases the threshold laws in quantum mechanics can be deduced
from general considerations without dynamical treatment.
For instance, the break-up cross section with $N$ fragments in
the final state and a short range interaction between them
could be estimated from simple phase-space volume (i.e. statistical)
arguments as
\begin{eqnarray}
\sigma_{\rm s} \sim E^{\frac{3}{2}(N-1) - 1} .
\end{eqnarray}
If one presumes that all fragments (``electrons'') are
{\it attracted}\/ by Coulomb forces to one fragment (``core''), but
the interaction between the ``electrons'' is negligible, then the
phase space arguments could be easily modified to give
\begin{eqnarray}
\sigma_{\rm C} \sim E^{N-2} .
\end{eqnarray}
In case of {\it repulsive}\/ Coulomb interaction with the ``core''
(but still without other interfragment interactions) the cross section
at the threshold becomes exponentially small, as obtained, for
example, by Geltman \cite{Gelt} in his calculations for atom
ionization by positron impact with all correlation neglected.
The threshold behaviour changes to $\sim E^{3/2}$ \cite{Sil} if one
employs the so called 3C wave functions for the final continuum state.
However, these functions do not ensure proper description in the
near-threshold domain.

If one aims to obtain a correct threshold law for the Coulomb
system, then the interaction between the fragments, i.e.
{\it the particle correlation}\/ is to be taken
into account. This makes the phase-space arguments insufficient, but
requires dynamical treatment as it was originally done by Wannier
\cite{Wannier} for the simplest system. In this paper we
employ the most
simple theoretical apparatus presenting the essential equations
in an arbitrary coordinate frame.
They remain valid, in particular,
in the simplest  single-particle
Cartesian coordinates.

As discussed in Section \ref{quant}, Poelstra {\it et al}\/ \cite{Poel}
suggested another formula for the Wannier index which differs
from our Eq.(\ref{ind}) by the extra ``phase factor'' $(N-2)$.
This discrepancy remains hidden when
one restricts consideration to the case of two, three, or four
electrons receding from the positively charged core.
In these cases the unstable mode proves to be respectively
non-degenerate, doubly- and triply-degenerate. Thus the degree of
degeneracy \underline{in these cases} coincides with $(N-2)$.
This fortuitously allows one to replace the summation over degenerate modes
implied by  formula (\ref{ind}) by multiplication over the factor
$(N-2)$ which corresponds to the formula by Poelstra
{\it et al} \cite{Poel}.
However this
coincidence is accidental and misleading. It is broken, for instance,
by variation of charges and masses of the constituent particles which
violates SCs symmetry and hence lifts the modes degeneracy, or by
considering larger numbers of particles $N$ (simply because
possible degrees of degeneracy are restricted by  properties
of the point groups in 3D space). For five electrons receding from a
charged core only doubly degenerate unstable modes were found above.

Physically it is clear that if  the charge of the core $Z$
in the system
${\rm A}^Z + (N-1) \, e$ becomes bigger, then
the interelectron correlations
should become less important and the threshold
law should approach
the value obtained from the phase-space arguments,
i.e. $\mu \rightarrow (N-2)$ as $Z \rightarrow \infty$. This
conclusion is supported by all examples considered. Moreover
in all these examples one can note that:

\begin{itemize}

\item
the number of
unstable modes accounting for their degeneracy  (i.e. the number
of terms in the sum (\ref{ind})) is equal to $(N-2)$;

\item
each partial Wannier index $\mu_{k 0}$ (\ref{partind}) tends to unity
from above as $Z$ increases.

\end{itemize}

Apparent exception from the first rule is an emergence of an additional
unstable mode in the plane $A^{+Z} + 4e$ SC for $Z=1$. However,
this SC provides $\mu$ larger than 3D SC and therefore it does not
govern the threshold behaviour.
Note that although these properties are physically very natural,
it is not clear if they can be proven rigorously from the
first principles.
An additional observation is that the electrons in SC tend to
be distributed uniformly on the sphere, even when the corresponding
perfectly symmetrical body does not exist (see five-electron case above).
For large number of electrons in the field of the core
several competing SC are found to produce very close
threshold indices. Still, in all the cases considered
the leading SC is found to be the three-dimensional one.

These results hopefully
should hold if the electrons are replaced by other
(possibly different) negatively charged particles. However, the situation
changes drastically if one of the ``electrons'' is replaced by a particle
of positive charge, for example, positron. It is essential that an
additional {\it repulsive Coulomb interaction}\/ appears in the system.
If  correlations are neglected then the cross section
decreases exponentially as $E$ approaches threshold. One could expect
that although the true threshold law retains a power character for all
values of $Z$, it tends to mock the exponential behaviour by
increasing of $\mu$ value \cite{ftnt}. This property holds for all
positron-containing systems considered above.
The threshold index increases with $Z$ quite slowly.
In order to illustrate the later point quantitatively we cite results
for ${\rm A}^{+Z} + 2e^- + e^+$ system with very large values of $Z$
(cf. Sec. \ref{parti}B):
$\mu =9.4$ for $Z=50$ ($\alpha = 5.60^\circ$, $\beta = 37.2^\circ$);
$\mu =11.6$ for $Z=100$ ($\alpha = 3.95^\circ$, $\beta = 37.8^\circ$).
In general terms one can argue that a similar situation
should arise when
a system contains two or more positively charged particles
{\it and}\/ two or more particles with negative charge.
Note that the properties of the partial Wannier indices  $\mu_{k 0}$
are less straightforward: some of them could be less than unity
and vary with $Z$ non-monotonically.

Large values of threshold indexes $\mu$ are unfavorable for an
experimental observation of the threshold behaviour: close to
the threshold the cross section proves to be too small to be observable,
and for higher excess energies the intrinsic deviations from the
threshold law become essential. An analysis of the energy domain
where the threshold law holds is beyond the scope of this paper.
Still, we can note that for the electron-impact ionization of atoms or
for double photoionization this domain is limited to few eV
above threshold (for quantitative treatment within the Wannier
mechanism see Refs.\cite{KO92} \cite{KOZPhys}). For the
positron-impact ionization the applicability domain is even
less \cite{KOZPhys} \cite{Posexp}. As argued by Ihra {\it et al}\/
\cite{Macekpos}, an agreement with experimental data could
be substantially improved if the interaction of different
modes in the deviation from SC is taken into account.
Possibly some procedure to assess for the mode interaction
could be developed also for the multifragment system; the
present development provides a  necessary first step for more
advanced approaches. One could note also that even very large
threshold indices could (quite unexpectedly) be useful
for constructing formulae of interpolation character as shown
in the recent paper by Rost and Pattard \cite{Rost}.

\acknowledgements

We are thankful to D.V.Dzuba for his help in numerical calculations
and to O.I.Tolstikhin for useful comments.
M.Yu.K. is appreciates support from the Australian Research Council.
V.N.O. acknowledges partial support from the
Russian Foundation for Basic Research under the grant 96--02--17023.
This work was supported by the Australian Bilateral Science and
Technology Collaboration Program.
V.N.O. is grateful for hospitality of the staff of the
School of Physics of UNSW where this work was carried out.

\appendix

\section{Eigenvalues of ${\bf V}$ matrix corresponding to rotations
and translations in time}

If the $N$-particle system is rotated as a whole over infinitisemal
(time-independent) angle $\delta \varphi$ around the axis $\vec{\nu}$,
then the particle coordinates receive increments
\begin{eqnarray} \label{rot}
\delta \vec{r}_j^{\, (\nu)} = (\vec{\nu} \times \vec{r}_j) \, \delta \varphi .
\end{eqnarray}
The form of Newtonian equations of motion
\begin{eqnarray}
m_j \, \frac{d^2 \vec{r}_j}{dt^2} =
- \frac{\partial U}{\partial \vec{r}_j}
\end{eqnarray}
remains invariant under rotations. This implies that
\begin{eqnarray} \label{inceq}
m_j \, \frac{d^2 \delta \vec{r}_j}{dt^2} =
- \sum_{i=1}^N \frac{\partial^2 U}{\partial \vec{r}_i \, \partial
\vec{r}_j} \, \delta \vec{r}_j ,
\end{eqnarray}
where $ \delta \vec r_j = \delta \vec r_j^{\, (\nu)}$.
For SC one can use Eqs.(\ref{rot}) and (\ref{scale})
to get
\begin{eqnarray} \label{der}
\frac{d^2 \delta \vec{r}_j^{\, (\nu)}}{dt^2} =
\frac{1}{\phi} \, \frac{d^2 \phi}{dt^2}\, \delta \vec{r}_j^{\, (\nu)}.
\end{eqnarray}
Bearing in mind that according to (\ref{Hamphi})
\begin{eqnarray} \label{d1}
\phi^2 \, \frac{d^2 \phi}{dt^2} = - \frac{{\cal Q}_0}{{\cal M}}
\end{eqnarray}
and using definition (\ref{V}), we finally obtain
\begin{eqnarray}
\frac{1}{m_j} \, \sum_{i=1}^N V_{ji} \, \delta \vec{r}_i^{\, (\nu)} =
\frac{{\cal Q}_0}{{\cal M}} \, \delta \vec{r}_j^{\, (\nu)}
\end{eqnarray}
which means that the grand vector ${\delta \bf r}^{\, (\nu)}$ is
an eigenvector of the grand matrix ${\bf K} \, {\bf V}$ with the eigenvalue
${\cal Q}_0/{\cal M}$. Generally there are three eigenvectors
corresponding to this eigenvalue, but for a linear SCs only two
independent rotations are possible.

Consider now variation of the trajectory
caused by shifting of time over an infinitesimal interval
$t\rightarrow t+\delta t$ using similar technique.
For the system in SC the
particle coordinates
are incremented in this case by
\begin{eqnarray} \label{prop}
\delta \vec{r}_j^{\, {\rm (SC)}} = \vec v_j \delta t = \frac{d \phi}{dt} \,
\vec{\rho}_j \, \delta t .
\end{eqnarray}
The form of Newtonian equations of motion obviously remains invariant
under the shift of the time variable. Therefore (\ref{inceq})
remains valid for
$ \delta \vec r_j= \delta \vec r ^{\, \rm (SC)}_j$.
An analogue of Eq.(\ref{der}) now reads
\begin{eqnarray}
\frac{d^2 \delta \vec{r}_j^{\, {\rm (SC)}}}{dt^2} =
\left( \frac{d \phi}{dt} \right)^{-1}
\frac{d^3 \phi}{dt^3}\, \delta \vec{r}_j^{{\rm \, (SC)}}.
\end{eqnarray}
Differentiating Eq.(\ref{d1}) we obtain
\begin{eqnarray}
\frac{d^3 \phi}{dt^3} = \frac{2 {\cal Q}_0}{{\cal M}} \,
\frac{1}{\phi^3} \, \frac{d \phi}{dt}
\end{eqnarray}
which  finally brings us to
\begin{eqnarray}
\frac{1}{m_j} \, \sum_{i=1}^N V_{ji} \, \delta \vec{r}_i^{\, {\rm (SC)}} =
- \frac{2{\cal Q}_0}{{\cal M}} \, \delta \vec{r}_j^{\, {\rm (SC)}} .
\end{eqnarray}
Since the grand vector ${\bf \delta r}^{\, {\rm (SC)}}$ is
proportional to the grand vector
${\mbox {\boldmath $\rho$} }$
which defines the SC shape,
we conclude that the latter vector is an eigenvector of the grand matrix
 ${\bf K} \, {\bf V}$ with the eigenvalue $- 2 {\cal Q}_0/{\cal M}$.


\begin{table} 
\caption{
Parameters of scaling configurations and Wannier indices for 
${\rm A}^{+Z} + 2e^- + e^+$ system.}
\begin{tabular}{c c c c c}
$Z$ & SC parameters & $\mu_1$ & $\mu_2$ & $\mu$ \\
\hline
\hline
\multicolumn{5}{c}
{SC $L_a$} \\
\hline
1 & $x=0.506100, \quad y=1.692952$ & 4.442178 & 2.193945 & 6.636123 \\
\hline
2 & $x=0.587468, \quad y=1.636629$ & 4.767141 & 2.064237 & 6.831377 \\
\hline
3 & $x=0.633155, \quad y=1.609587$ & 5.024502 & 1.966884 & 6.991386 \\
\hline
4 & $x=0.664214, \quad y=1.594313$ & 5.242328 & 1.88483  & 7.127158  \\
\hline
\hline
\multicolumn{5}{c}
{SC $L_b$} \\
\hline
1 & $x=0.441380, \quad y=-0.677611$ & 2.577720 & 1.025435 & 3.603155 \\
\hline 
2 & $x=0.539724, \quad y=-0.847969$ & 2.888492 & 1.009213 & 3.897705 \\
\hline
3 & $x=0.594480, \quad y=-0.949091$ & 3.193559 & 1.005040 & 4.198599 \\
\hline
4 & $x=0.631720, \quad y=-1.023071$ & 3.475766 & 1.003244  & 4.479010 \\
\hline
\hline
\multicolumn{5}{c}
{SC $P$} \\
\hline
1 & $2 \alpha = 76.7338^\circ$, \quad $2 \beta = 55.1969^\circ$ & 
1.884950 & 1.562234 & 3.447184 \\
\hline
2 & $2 \alpha = 55.1741^\circ$, \quad $2 \beta = 61.3793^\circ$ & 
2.045028 & 1.793101 & 3.838128 \\
\hline
3 & $2 \alpha = 45.4233^\circ$, \quad $2 \beta = 64.1787^\circ$ & 
2.206553 & 1.972092 & 4.178645 \\
\hline
4 & $2 \alpha = 39.5138^\circ$, \quad $2 \beta = 65.8916^\circ$ & 
2.351217 & 2.123469 & 4.474686 \\
\end{tabular}
\end{table}

\begin{table} 
\caption{Wannier indices for  ${\rm A}^{+Z} + 4e$ system.
The numbers in parentheses indicate degree of unstable mode 
degeneracy.}
\begin{tabular}{c l l l c}
$Z$ & $\mu_1$ & $\mu_2$ & $\mu_3$ & $\mu$ \\
\hline
\hline
\multicolumn{5}{c}
{SC $P$} \\
\hline
1 & 4.877419 & 4.248225 (2) & 2.071837 & 15.44571 \\
\hline
2 & 1.356093 & 1.273381 (2) & --- & 3.902855 \\
\hline
3 & 1.192808 & 1.145660 (2) & --- & 3.484128 \\
\hline
4 & 1.132414 & 1.099316 (2) & --- & 3.331046 \\
\hline
5 & 1.100871 (2) & 1.075346 & --- & 3.251563 \\
\hline
\hline
\multicolumn{5}{c}
{SC $V$} \\
\hline
1 & 3.075960 (3) & --- & --- & 9.227870  \\
\hline
2 & 1.257986 (3) & --- & --- & 3.773958  \\
\hline
3 & 1.139795 (3) & --- & --- & 3.419384  \\
\hline
4 & 1.095940 (3) & --- & --- & 3.287819  \\
\hline
5 & 1.073040 (3) & --- & --- & 3.219120   \\
\end{tabular}
\end{table}

\newpage
\begin{table} 
\caption{
Parameters of SCs and Wannier indices for 
${\rm A}^{+Z} + 3e^- + e^+$ system. 
The numbers in parentheses indicate degree of unstable mode degeneracy.}
\begin{tabular}{c c c c c c}
$Z$ & SC parameters & $\mu_1$ & $\mu_2$ & $\mu_3$ & $\mu$ \\
\hline
\hline
\multicolumn{6}{c}
{SC $L$} \\
\hline
1 & $x=0.580448$, \quad $y=-1.070391$ \quad $z =1.627861$ & 
4.41213 & 2.30976 & 1.06890 & 7.79079 \\
\hline
2 & $x=0.580448$, \quad $y=-1.070391$ \quad $z =1.627861$ & 
4.73844 & 2.13221 & 1.03343 & 7.90408 \\
\hline
3 & $x=0.628772$, \quad $y=-1.162883$ \quad $z =1.602043$ & 
4.99954 & 2.01766 & 1.02270 & 8.03998 \\
\hline
4 & $x=0.661096$, \quad $y=-1.22438$ \quad $z =1.587485$ & 
5.21972 & 1.92598 & 1.01746 & 8.16316 \\
\hline
\hline
\multicolumn{6}{c}
{SC $V$} \\
\hline
1 & $\alpha = 60.5698^\circ$, \quad $\beta = 32.2041^\circ$ & 
1.57584 (2) & 1.03194 & 0.60493 (2) & 5.39348 \\
\hline
2 & $\alpha = 40.5400^\circ$, \quad $\beta = 41.7154^\circ$ & 
1.56354 (2) & 1.20043 & 0.66302 (2) & 5.65356 \\
\hline
3 & $ \alpha = 32.3675^\circ$, \quad $\beta = 44.9869^\circ$ & 
1.70957 (2) & 1.33711 & 0.62771 (2) & 6.01166 \\
\hline
4 & $ \alpha = 27.6668^\circ$, \quad $\beta = 46.7663^\circ$ & 
1.85129 (2) & 1.45098 & 0.57327 (2) & 6.30011 \\
\end{tabular}
\end{table}

\begin{table} 
\caption{
Parameters of SCs and Wannier indices for ${\rm A}^{+Z} + 5e$ system.
The numbers in parentheses indicate degree of unstable mode degeneracy.
}
\begin{tabular}{c c c c c c}
$Z$ & SC parameters & $\mu_1$ & $\mu_2$ & $\mu_3$ & $\mu$ \\
\hline
\hline
\multicolumn{5}{c}
{SC $P$} \\
\hline
\hline
2 & --- & 1.818250 (2) & 1.575289 (2) & 0.701595 (2) & 6.787079 \\
\hline
3 & --- & 1.363938 (2) & 1.245279 (2) & --- & 5.218433 \\
\hline
4 & --- & 1.235701 (2) & 1.156156 (2) & --- & 4.783715 \\
\hline
5 & --- & 1.174520 (2) & 1.114540 (2) & --- & 4.578120 \\
\hline
6 & --- & 1.138614 (2) & 1.090432 (2) & --- & 4.458093\\
\hline
7 & --- & 1.114982 (2) & 1.074705 (2) & --- & 4.379373 \\
\hline
\hline
\multicolumn{5}{c}
{SC $V$} \\
\hline
\hline
2 & $\alpha=45.15762^\circ$ & 1.606923 & 1.504688 & 1.493106 (2) & 6.097823 \\
\hline
3 & $\alpha=45.09672^\circ$ & 1.280163 & 1.228075 & 1.223717 (2) & 4.955672 \\
\hline
4 & $\alpha=45.06976^\circ$ & 1.182908 & 1.147576 & 1.145087 (2) & 4.620659 \\
\hline
5 & $\alpha=45.05455^\circ$ & 1.135887 & 1.109091 & 1.107406 (2) & 4.459790 \\
\hline
6 & $\alpha=45.04479^\circ$ & 1.108127 & 1.086528 & 1.085274 (2) & 4.365202 \\
\hline
7 & $\alpha=45.03799^\circ$ & 1.089794 & 1.071698 & 1.070708 (2) & 4.302909 
\\
\end{tabular}
\end{table}

\begin{table} 
\caption{Parameters of SCs and Wannier indices for 
$2 \, X^Z_m + 2e$ system.
The numbers in parentheses indicate degree of unstable mode degeneracy.}
\begin{tabular}{c l l l l l r}
$Z$ & $m$ & $\alpha$ & $\mu_1$ & $\mu_2$ & $\mu_3$ & $\mu$ \\
\hline
\hline
1 & 1 & $\alpha = 45^\circ$ & 1.29366 & 0.90584 (2) & --- & 3.10533 \\
\hline
2 & 1 & $\alpha = 32.2093^\circ$ & 1.36762 & 1.33643 & --- & 2.70405 \\
\hline
1 & 2 & $ \alpha = 35.9490^\circ$ & 1.56958 & 1.31788 & 0.54315 & 3.43062 \\
\hline
1 & 1836 & $ \alpha = 30.0049^\circ$ & 50.32979 & 37.46232 & --- & 87.79211 
\end{tabular}
\end{table}


\begin{references}
\bibitem{ftnt1}
Some important details of the derivation were omitted, especially
expression of the interaction potential in terms of new coordinates.
The hyperspherical approach was not applied to multifragment positron
containing systems.
\bibitem{angmod}
Note that this reasoning applies only to unstable modes. For the
stable modes (which are responsible for particle angular correlation)
the situation is drastically different, as discussed in detail
in Ref.\cite{KOWang}; see also Ref.\cite{EWRM}.
\bibitem{f1}
It is worth mentioning that in variance with Klar and
Schlecht we obtain correct value of $\mu$ without appealing to
the contributions from undetermined angular variables
$\psi$ and $\varphi$, see p.1709 of Ref.\cite{KS}.
\bibitem{GruFizCom}
The SCs $L_b$ and $P$ for this system were discussed originally
by Gruji\'{c} \cite{Gru2ep}. However the configuration $P$ was
discarded and the threshold indices evaluated for $L_b$ differ
from these reported in the present paper.
\bibitem{ftnt}
For the particular case of three-particle system the increase
of $\mu$ with $Z$ was noticed by Klar \cite{Klarpos}.
\end{references}
\end{document}